\title{Composite Likelihood Inference  \\ by Nonparametric Saddlepoint Tests}
\date{May 2013}

\documentclass[12pt,a4paper,twoside]{article}
\usepackage[utf8]{inputenc}
\usepackage{amsmath}
\usepackage{bbm}
\usepackage{booktabs}
\usepackage{natbib}
\usepackage{hyperref}
\usepackage{subfigure}
\usepackage{graphicx}
\usepackage{fullpage}
\usepackage{authblk}

\author{Nicola Lunardon}
\affil{Department of Economics, Business, Mathematics and Statistics ``Bruno de Finetti'', University of Trieste, Piazzale Europa 1, 34127, Trieste, Italy\\nicola.lunardon@econ.units.it}

\author{Elvezio Ronchetti}
\affil{Research Center for Statistics and Dept. of Economics, \authorcr University of Geneva, 1211, Geneva, Switzerland\\elvezio.ronchetti@unige.ch}

\newcommand\solidrule[1][.2cm]{\rule[0.5ex]{#1}{.9pt}}
\newcommand\dashedrule{\mbox{%
  \solidrule[1.3mm]\hspace{1.5mm}\solidrule[1.3mm]}}
\newcommand\dottedrule{\mbox{%
  \solidrule[.4mm]\hspace{1mm}\solidrule[.4mm]\hspace{1mm}\solidrule[.4mm]}}
\newcommand\dotteddashrule{\mbox{%
  \solidrule[.4mm]\hspace{1mm}\solidrule[1.6mm]\hspace{1mm}\solidrule[.4mm]}}
\newcommand\longdashedrule{\mbox{%
  \solidrule[1.8mm]\hspace{1.5mm}\solidrule[1.8mm]}}
\newcommand\shortlongdashedrule{\mbox{%
  \solidrule[.8mm]\hspace{1mm}\solidrule[1.8mm]\hspace{1mm}\solidrule[.8mm]}}

\def\T{{ \mathrm{\scriptscriptstyle T} }}

\begin{document}
\maketitle
\newpage

\begin{abstract}
The class of composite likelihood functions provides a flexible and powerful toolkit to carry out approximate inference for complex statistical models when the full likelihood is either impossible to specify or unfeasible to compute.
However, the strenght of the composite likelihood approach is dimmed when considering hypothesis testing about a multidimensional parameter because the finite sample behavior of likelihood ratio, Wald, and score-type test statistics is tied to the Godambe information matrix. Consequently inaccurate estimates of the Godambe information translate in inaccurate $p$-values. 
In this paper it is shown how accurate inference can be obtained by using a fully nonparametric saddlepoint test statistic derived from the composite score functions. The proposed statistic is asymptotically chi-square distributed up to a relative error of second order and does not depend on the Godambe information. The validity of the method is demonstrated through simulation studies.
\end{abstract}


\newpage

\section{Introduction}
The likelihood function plays a central role in statistical inference.
However, with statistical models becoming increasingly complex in many fields such as genetics and
finance, the full likelihood function is often not available in closed form or is too difficult to
specify. This can be due for instance to a complex dependence structure of the data. Examples
include e.g. the estimation of diffusion models in finance and models based on
max-stable processes for spatial multivariate extremes (\citet{padoan10},
\citet{Thibaud13}).
Even when the specification of the full likelihood is straightforward, its evaluation can be computationally awkward. For instance, modeling a spatial process with a Gaussian random field requires the determinant and the inverse of the process covariance matrix, whose dimension grows as the number of observed sites increases
\citep{Stein2004}.

In these cases and in the frequentist setting, one can rely on indirect inference techniques
(see the surveys by \citet{HegglandFrigessi2004} and \citet{JiangTurnbull2004}), whereas in the Bayesian framework one can use sequential Monte Carlo methods for approximate Bayesian computations (see, for instance \citet{DelMoral2006}, \citet{Beaumont2009}). 

An attractive alternative which has gained popularity in the past few years is the approach based on
composite likelihood functions originally proposed by \cite{Lindsay}. The basic idea is to approximate
the unknown full likelihood by a sum of likelihood components obtained e.g. by combining either marginal or conditional densities.
An important special case is the pairwise likelihood constructed using
pairs of components; see \cite{CoxReid}. Although the resulting combined function is no longer a proper
likelihood, the derived
inferential procedures are $M$-estimators and tests based on unbiased estimating functions. From a
theoretical point of view this is an appealing property because their asymptotic theory is readily available;
cf. e.g. \cite{HeritierRonchetti1994} in the context of robust tests. 
Specifically, Wald and score test statistics for pairwise likelihoods are asymptotically
$\chi^2$ distributed, whereas the asymptotic distribution of the pairwise log-likelihood ratio test
statistic is a linear combination of independent $\chi_1^2$ random variables.   

The use of composite likelihoods has been advocated by several authors both in the frequentist setting
(see the good review paper by \cite{VarinReidFirth2011} in a special issue devoted to this topic in
{\it Statistica Sinica}) and also in the Bayesian framework \citep{PauliRacugnoVentura2011, Ribatet11}. 
Successful use of this approach in fairly complex models include applications in spatial processes
(\cite{HeagertyLele1998}, \cite{VarinHostSkare2005}), generalized linear mixed models
(\cite{RenardMolenberghsGeys2004}, \cite{BellioVarin2005}), longitudinal models
\citep{FieuwsVerbeke2006}, and genetics (\cite{Hudson2011}, \cite{McVean2004}).

In spite of the availability of standard asymptotic theory for Wald, score, and likelihood ratio tests
based on pairwise likelihoods, their actual computation requires the evaluation of the expectations
of minus the derivative and of the square of the pairwise likelihood score which,
as opposite to the full likelihood score, are not equal.
Their estimation in this case is akward and the corresponding $p$-values and coverage probabilities
based on the asymptotic distribution
become inaccurate when the sample size is moderate or when small tail probabilities are
required; cf. Section \ref{pairwise} and \ref{examples}.
To improve the accuracy, the test statistics could be adjusted as in the classical case
by means of Barlett corrections and related methods. However, these methods would provide only
improvements in terms of the absolute error of the approximation which would still be inaccurate in
the tails. 

In this paper we consider an alternative test for pairwise likelihood defined by (\ref{pwsp}).
It is a nonparametric test derived by building on the results by
\cite{RobinsonRonchettiYoung2003}.
It enjoys the following desirable properties: i) the test statistic is asymptotically $\chi^2$ distributed; ii) the $\chi^2$ approximation to the exact distribution has a {\it relative error} of order $O(n^{-1})$; iii) the test is fully nonparametric; iv) the test can combine accuracy and robustness by an appropriate choice of the pairwise likelihood score; v) the test does not require the computation of elements of the asymptotic covariance matrix of $M$-estimators (so-called sandwich formula or Godambe information); vi) the test statistic is parametrization invariant.


These properties will be discussed in detail in Section \ref{saddlepoint} and make this
test an attractive alternative for inference with pairwise likelihoods.

The rest of the paper is organized as follows.
In Section \ref{pairwise} we define the pairwise likelihood and discuss the available
test procedures. In Section \ref{saddlepoint} we introduce the new test and discuss its
properties. Section \ref{examples} present three examples that show the excellent finite
sample behavior of the new test. Finally, some conluding remarks and an outlook are
given in Section \ref{conclusions}.

\section{Pairwise Likelihood}\label{pairwise}
Let $y=(y_1,\dots, y_n)^\T,$ be a random sample of independent realizations of the $q$-dimensional random vector $Y$ having probability distribution $F(\cdot;\theta)$ and density function $f(\cdot;\theta),\,\theta\subseteq \mathbbm R^p$. The full log-likelihood function and ratio are respectively $\ell(\theta)=\log f(y;\theta)$ and $w(\theta)=2[\ell(\hat\theta)-\ell(\theta)]$, with $\hat\theta$ the maximum likelihood estimate. 
Consider a set of measurable events $\left\{\mathcal E_r\in\mathcal Y,\,r=1,\dots,m\right\}$ on the sample space $\mathcal Y$, defined for pairs of components $(y_{ij},y_{ik}),\,j\neq k=1,\dots,q$, and let $f_r(y; \theta)=f(y\in\mathcal E_r; \theta)$ be the likelihood contribution generated from $f(y;\theta)$ by considering the event $\mathcal E_r$. Then the pairwise log-likelihood is defined as 
\begin{equation}
\label{pwl}
	p\ell(\theta)=\sum_{i=1}^n\sum_{r=1}^m \omega_{ir}\log f_r(y_i; \theta),
\end{equation}
where $\omega_{ir}$ are weights not depending on $\theta$ nor $y$. In general these weights are chosen both to improve the efficiency of the maximum pairwise likelihood estimator and to reduce the computational effort \citep{LinYiSun}. 
The pairwise score function associated to \eqref{pwl} is
\begin{equation*}
\label{pws}
	ps(\theta)=\sum_{i=1}^n\sum_{r=1}^m \omega_{ir}\frac{\partial\log f_r(y_i; \theta)}{\partial\theta}=\sum_{i=1}^n ps(\theta; y_i).
\end{equation*}
Since it is a combination of genuine scores, $ps(\theta)$ is an unbiased estimating function, that is $ \mathbbm E_F\left[ ps(\theta) \right ]=0$, where the notation $ \mathbbm E_F$ is used to highlight that expectation is taken with respect to the full model.

The maximum pairwise likelihood estimator $\hat\theta_p$ belongs to the class of M-estimators and is implicitly defined through the equation 
\begin{equation*}
\label{pws_estimating_eq}
ps(\theta)=0.	
\end{equation*}

Under broad conditions \citep[see, e.g.,][]{MoleVerb}, the maximum pairwise likelihood estimator is consistent and asymptotically normal, with covariance matrix given by the so-called sandwich formula or
expected Godambe information
\begin{equation*}
\label{godambe}
	V(\theta)=H(\theta)^{-1}J(\theta)H(\theta)^{-1},
\end{equation*}
where $J(\theta)= \mathbbm E_F\left[ ps(\theta;Y)ps(\theta;Y)^\T \right]$, $H(\theta)=-\mathbbm E_F\left[\partial ps(\theta;Y)/\partial\theta^\T \right]$. 

In the context of hypothesis testing, the pairwise likelihoods allow to perform the analogous of the Wald, the score and the likelihood ratio tests. The pairwise likelihood counterparts of the Wald and score test statistics are 
\begin{equation*}
\label{tests_WS}
pw_w(\theta)=n(\hat\theta_p-\theta)^\T V(\hat\theta_p)^{-1}(\hat\theta_p-\theta)\quad\text{and}\quad pw_s(\theta)=n^{-1}ps(\theta)^\T J(\theta)^{-1} ps(\theta),
\end{equation*}
respectively. Under the hypothesis $H_0:\theta=\theta_0$ both $pw_w(\theta_0)$ and $pw_s(\theta_0)$
converge to a chi-square distribution with $p$ degrees of freedom.  
Instead, the pairwise log-likelihood ratio 
\begin{equation*}
\label{pwr}
	pw(\theta)=2\left\{ p\ell(\hat\theta_p)-p\ell(\theta) \right\}
\end{equation*}
converges in distribution to $\sum_{j=1}^p \lambda_j(\theta)Z^2_j$, where $\lambda_1(\theta),\dots,\lambda_p(\theta)$
are the eigenvalues of $H(\theta)^{-1}J(\theta)$ and the $Z_j's$ independent random variables with a standard normal distribution \citep[see, e.g.,][]{Kent1982}. 
Adjustments to $pw(\theta)$ have been proposed to provide a pairwise log-likelihood ratio with the usual asymptotic chi-square distribution. The simplest adjustment is based on first moment matching 
\begin{equation*}
\label{pw1}
	pw_1(\theta)=\frac{pw(\theta)}{\kappa_1},
\end{equation*}
where $\kappa_1= \mathbbm E\left[ \sum_{j=1}^p \lambda_j(\theta)Z^2_j \right]/p= \sum_{j=1}^p\lambda_j(\theta)/p$. A $\chi^2_p$
approximation is used for the distribution of $pw_1(\theta)$ \citep[see, e.g.][]{RotJew}.  
Alternatively, \citet{ChanBate} propose the so-called vertical scaling to $pw(\theta)$
\begin{equation}
\label{cb}
	pw_{cb}(\theta)=\frac{pw_w(\theta)}{\kappa_{cb}},
\end{equation}
where $\kappa_{cb}=n(\hat\theta_p-\theta)^\T H(\hat\theta_p)(\hat\theta_p-\theta)/pw(\theta)$.
Finally, \citet{PSS} propose a parametrization invariant adjustment 
\begin{equation}
\label{pss}
	pw_{inv}(\theta)=\frac{pw_s(\theta)}{\kappa_{inv}},
\end{equation}
where $\kappa_{inv}=n^{-1} ps(\theta)^\T H(\theta)^{-1}ps(\theta)/pw(\theta)$. Test statistics \eqref{cb} and \eqref{pss} are first order equivalent to $pw_w(\theta)$ and $pw_s(\theta)$ respectively and are
asymptotically $\chi^2_p$ distributed. 
Even with these adjustments, the $\chi^2$ approximation for the distribution of these test statistics may be inaccurate in moderate sample sizes or when small tail probabilities are required. The accuracy of the approximation mostly depends on the Godambe information matrix, as can be seen from the definition of the test statistics. To better understand this statement it is important to distinguish two relevant settings in the pairwise likelihood framework. In the first one, pairwise likelihoods replace the full likelihood function for computational convenience. Therefore, either analytic expressions or (parametric) bootstrap estimates for $J(\theta)$ and $H(\theta)$ can be worked out under the assumed $F(;\theta)$. In the second one, pairwise likelihoods are used as an approximation to $\ell(\theta)$ and in this case only empirical counterparts of such matrices can be computed. In the case of independent observations the estimates  
\begin{equation*}
\label{est_JH}
	\hat J(\theta)=\frac{1}{n} \sum_{i=1}^n ps(\theta; y_i)ps(\theta; y_i)^\T\quad\text{and}\quad \hat H(\theta)=-\frac{1}{n}\sum_{i=1}^n \frac{\partial ps(\theta; y_i) }{ \partial\theta^\T },
\end{equation*}
are consistent for $J(\theta)$ and $H(\theta)$, respectively. However, depending on the application area, $\hat J(\theta)$ may not be appropriate and a consistent estimate should be obtained by using resampling methods \citep[see][and references therein]{VarinReidFirth2011}.

The second setting is the most likely to occur in real applications and it is the most critical. Indeed, the estimation of $J(\theta)$ and $H(\theta)$ introduces additional variability and deteriorates the accuracy of the $\chi^2$ approximation in finite samples. In the next section we present an alternative test which avoids these problems.

\section{Saddlepoint Test}\label{saddlepoint}
Consider for simplicity of notation the case of a simple hypothesis.
The new test statistic is

\begin{equation}
\label{pwsp}
pw_{sp}(\theta)=
 -2n \log \Big \{\sum_{i=1}^n w_i(\theta) \exp \{\lambda(\hat\theta_p)^\T ps(\hat\theta_p;y_i) \} \Big \} ,
\end{equation}
where
\begin{equation*} 
\label{w_i}
w_i(\theta)= \exp \{\beta(\theta)^\T ps(\theta;y_i) \}/
             \sum_{j=1}^n \exp \{\beta(\theta)^\T ps(\theta;y_j) \} ,
\end{equation*}
$\beta(\theta)$ is the root of the equation
\begin{equation}
\label{beta_theta}
\sum_{i=1}^n w_i(\theta) ps(\theta;y_i) = 0 ,
\end{equation}
and 
$\lambda(\hat\theta_p)$ satisfies the equation
\begin{equation*}
\label{lambda_theta_p}
\sum_{i=1}^n ps(\hat\theta_p;y_i) \exp \{\lambda(\hat\theta_p)^\T ps(\hat\theta_p;y_i) \}=0.
\end{equation*}

The following theorem states the large sample properties of $p$-values obtained from test statistic \eqref{pwsp}. The proof is provided in the Appendix.

{\bf Theorem}. 
\emph{
Suppose that conditions (A.1), (A.2), and (A.3) in the Appendix hold. Then under the null hypothesis $H_0:\theta=\theta_0$
\begin{displaymath}
\label{asymptotic_distribution_pwsp}
P_{H_0}[pw_{sp}(\theta_0)\geq pw_{sp}(\theta_0)^{obs}]=(1-Q_p(pw_{sp}(\theta_0)^{obs}))(1+O_p(n^{-1}))
\end{displaymath} 
where $pw_{sp}(\theta_0)^{obs}$ is the observed value of the statistic and $Q_p(\cdot)$ is the distribution function
of a chi-square random variable with $p$ degrees of freedom.
}


The test statistic (\ref{pwsp}) can be rewritten as
$pw_{sp}(\theta)= -2n \hat K_w(\lambda(\hat\theta_p), \hat\theta_p),$
where $\hat K_w(\cdot; \cdot)$ is the cumulant generating function of $ps(\cdot;Y)$ under
the discrete distribution defined by $\{w_i\}$, with $w_i=w_i(\theta)$. The latter is the discrete distribution which
is closest to the empirical one $\{ \frac{1}{n} \}$ with respect to the {\it backward} Kullback-
Leibler divergence 
\begin{equation*}
\label{backwardKL}
d_{KL}(\{w_i\},\{\frac{1}{n}\}) = 
\sum_{i=1}^n w_i \log \Big [\frac{w_i}{1/n} \Big ] =
\sum_{i=1}^n w_i \log w_i + \log n
\end{equation*}
and which makes $ps(\theta)$ unbiased (see equation (\ref{beta_theta})). Notice that
the use of the {\it forward} Kullback-Leibler divergence
\begin{equation*}
\label{forwardKL}
d_{KL}(\{\frac{1}{n}\},\{w_i\}) = 
\sum_{i=1}^n \frac{1}{n} \log \Big [\frac{1/n}{w_i} \Big ] =
- \frac{1}{n} \sum_{i=1}^n \log w_i - \log n
\end{equation*}
would lead to the
classical empirical log-likelihood ratio test statistic \citep{Owen2001} which is also
asymptotically $\chi^2_p$ distributed,
but which does not enjoy the second-order relative error property of the present test.

Let us now discuss in more details the properties of this test which are
summarized in the Introduction.

The new test statistic is asymptotically $\chi^2$ distributed, therefore it is, up to first-order, equivalent to the standard tests but it differs for the following relevant features. Firstly, $pw_{sp}(\theta)$ is asymptotically pivotal and the result does not depend on suitable scaling factors, contrasted to the approximate pivots proposed by \citet{RotJew}, \citet{ChanBate}, and \citet{PSS}. Secondly, as $pw_{sp}(\theta)$ stems from a small sample asymptotics framework, it introduces an unexplored stream in the pairwise likelihood setting concerning the accuracy of tests statistics. In particular, the exact distribution of our test proposal is $\chi^2$ up to a relative error of magnitude $O(n^{-1})$. This provides an excellent accuracy uniformly in the tails for the approximation obtained by using the asymptotic distribution. 
Thirdly, the asymptotic approximation can not be enhanced by bootstrap calibration as the actual distribution of $pw_{sp}(\theta)$ and its bootstrap counterpart $pw_{sp}^{*}(\theta)$ are also distant by a relative error of order $O(n^{-1})$. In contrast, resorting to a computationally expensive resampling procedure is the only viable path either to estimate the quantiles of $pw(\theta)$ without computing the elements of the Godambe information \citep[see, e.g.][]{aerts01} or to obtain refined estimates of $J(\theta)$ and $H(\theta)$ \citep[see][Section 5.1]{VarinReidFirth2011}.
Fourtly, the test is fully nonparametric and depends only on the function $ps(\theta;y)$. Therefore, it does not require the specification of the full model $F(\cdot;\theta)$ which is clearly a key issue in this setup (see Section~\ref{pairwise}). Furthermore, as it solely depends on $ps(\theta;y)$, by choosing the latter bounded with respect to $y$ we can combine accuracy in small samples and resistance with respect to potential outliers; see \citep{LoRonchetti2012} in the GMM framework and the second example in Section \ref{examples} below.
Finally, $pw_{sp}(\theta)$ enjoys the desirable property of invariance under reparametrization as well as $pw(\theta)$, $pw_s(\theta)$, and $pw_{inv}(\theta)$. However, the latter lose exact invariance once the empirical estimates $\hat J(\hat\theta_p)$ and $\hat H(\hat\theta_p)$ are used.

\section{Numerical Examples}\label{examples}
This section aims at showing some numerical evidence about the behaviour of the nonparametric saddlepoint test statistic in the pairwise likelihood framework. Three examples will be illustrated, each of them enlightening a different feature of the test.

In the first example, the new test is compared to the pairwise likelihood ones presented in Section~\ref{pairwise}, and their finite sample accuracy to the $\chi^2$ approximations is analized in the context of a multivariate normal model.


In the second and third example, we consider a first-order autoregressive and a geostatistical model, respectively. 
The purpose of these examples is twofold. In first place we want to point out that the use of bounded estimating functions to compute $pw_{sp}(\theta)$ is recommended not only to provide versions of $pw_{sp}(\theta)$ whose accuracy remains stable under contaminations of the model.
Indeed, we will provide empirical evidence that supports, in this setup, the following results outlined in the Appendix: a) $pw_{sp}(\theta)$ converges to the $\chi^2$ distribution and the approximation has a relative error of second order; b) a second order agreement also holds between the asymptotic distribution of $pw_{sp}(\theta)$ and its bootstrap distribution $pw_{sp}^{*}(\theta)$.
In second place, these models provide a challenging setting in which $n=1\ll q$ and consequently a suitable definition of the pairwise likelihood function is needed.

In the first two examples the full log-likelihood function $\ell(\theta)$ is available and this allows us to set the log-likelihood ratio test $w(\theta)$ as a benchmark. In the third example this is not possible because the evaluation of the likelihood function is computationally prohibitive.

The statistical environment R \citep{R} was used to carry out all the computations in this paper.


\subsection{Multivariate Normal Model}
Let $Y$ be a normally distributed random vector, with expectation $(\mu,\dots,\mu)^\T\in \mathbbm R^q$ and covariance matrix $\Sigma$ having diagonal elements $\sigma^2$ and off-diagonal ones $\sigma^2\rho$, $\rho\in(-1/(q-1),1)$. The pairwise log-likelihood for the parameter $\theta=(\mu, \sigma^2, \rho)$ is
\begin{equation*}
\label{PL_ex1}
pl(\theta)=-\frac{nq(q-1)}{2}\left[\log\sigma^2+\frac{\log(1-\rho^2)}{2}\right]-\frac{1}{2\sigma^2(1-\rho^2)}\sum_i (y_i-\mu)^\T\Gamma(\theta)(y_i-\mu),
\end{equation*}
with $y_{i\cdot}=\sum_j y_{ij}$, $\Gamma_{jj}(\theta)=(q-1)$, $\Gamma_{jk}(\theta)=-\rho$, $j\neq k=1,\dots,q$. 

We run simulations by generating 100000 samples of size $n=10$ from $Y\in \mathbbm R^{30}$, with $\mu=0$, $\sigma^2=1$, and $\rho$ ranging from moderate to strong correlation values.

For each sample we computed the nonparametric saddlepoint statistic as well as those discussed in Section~\ref{pairwise}.
As for this example, $J(\theta)\,\text{and}\,H(\theta)$ are available \citep[see,][]{PSS}, this allows us to compare also the finite sample behavior among pairwise likelihood test statistics computed by using the exact matrices and their empirical counterparts $\hat J(\theta)\,\text{and}\, \hat H(\theta)$. In the following, the superscript $e$ will refer to statistics evaluated using $J(\theta)\,\text{and}\,H(\theta)$.

Table~\ref{tab:TAB1} reports empirical coverage probabilities for three dimensional confidence regions for $\theta$. As expected, the best results are obtained when the elements of the expected Godambe information are used and, in particular, when one considers $pw_s^e(\theta)$ and $pw^e_{inv}(\theta)$. However, it should be stressed that, in most real applications, only the observed Godambe information is available. In this case, pairwise likelihood statistics have empirical coverages far from the nominal levels. Instead, the bootstrap distribution of the nonparametric saddlepoint test statistic $pw_{sp}^{*}(\theta)$ is approximated quite well by the $\chi^2_3$ and the approximation is close to the one provided by the gold standard $w(\theta)$.
From simulation studies (not reported here) it is shown that confidence sets based on pairwise likelihood statistics achieve the nominal levels either by increasing the sample size or by using resampling-based estimates of $J(\theta)$ and $H(\theta)$.

\begin{figure}[!h]
\begin{center}
	\begin{tabular}{ccc}
	\quad\;{\bf {\small(a)} } & \quad\;{\bf {\small(b)} } & \quad\;{\bf {\small(c)} } \vspace{.2cm} \\
	\includegraphics[scale=0.27]{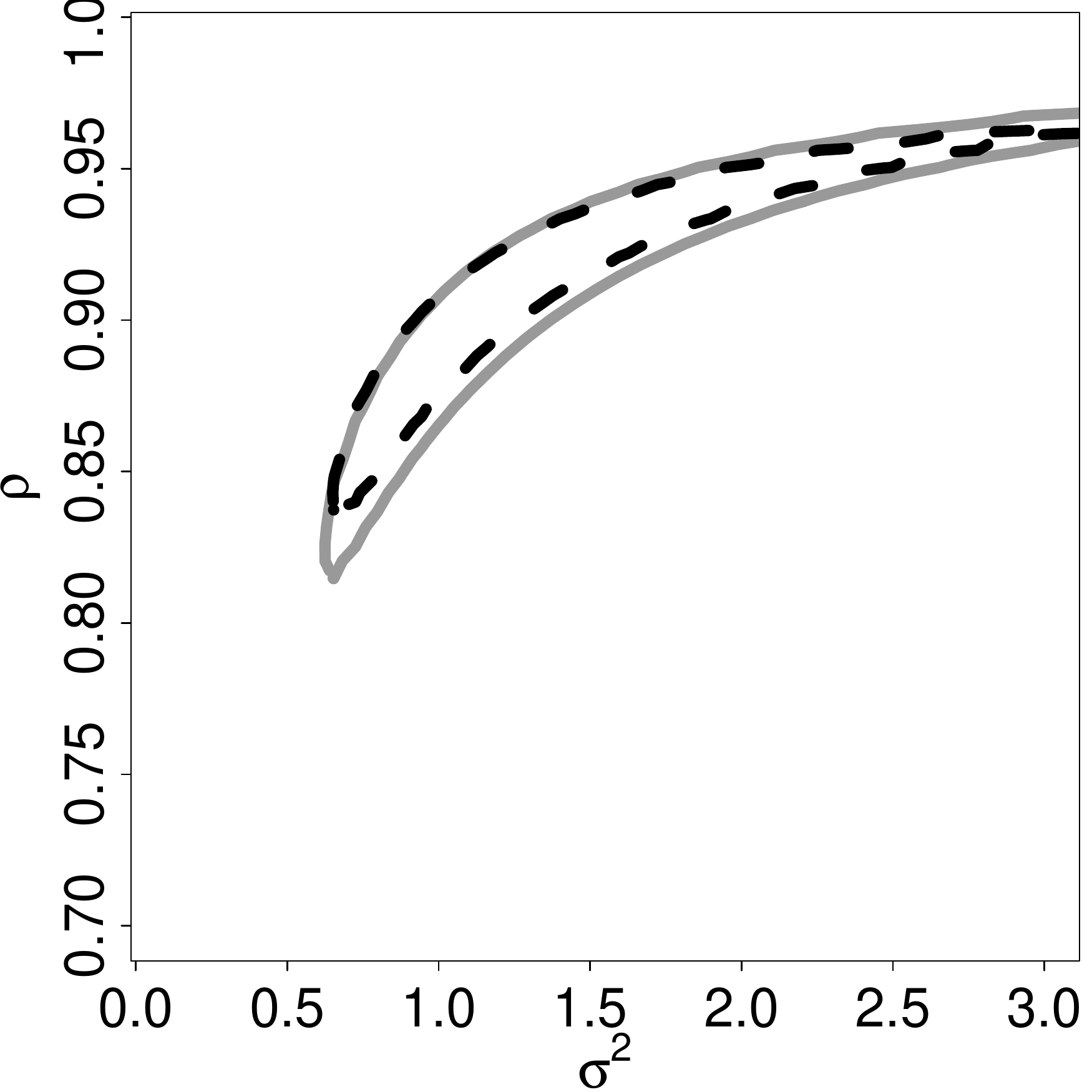} & \includegraphics[scale=0.27]{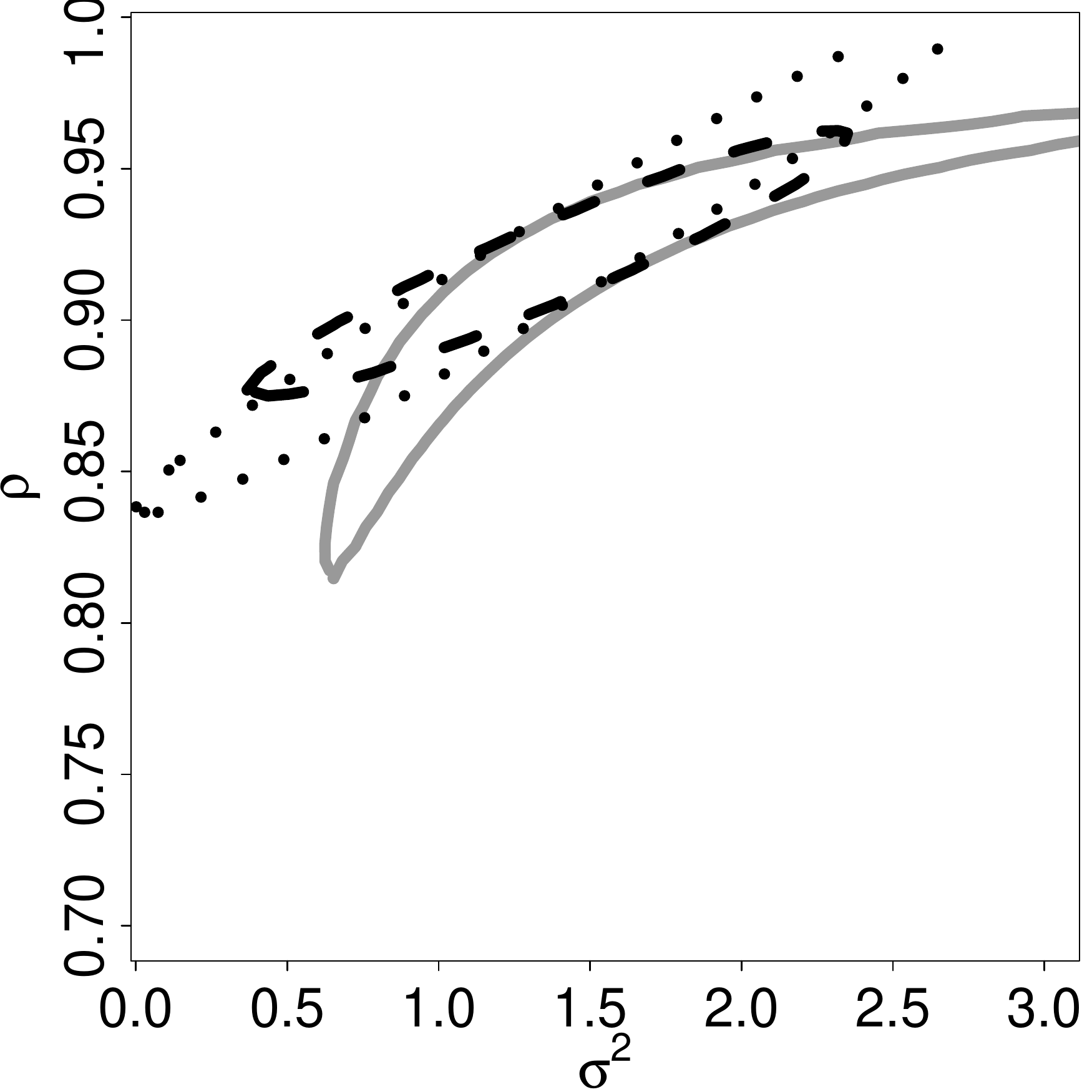} & \includegraphics[scale=0.27]{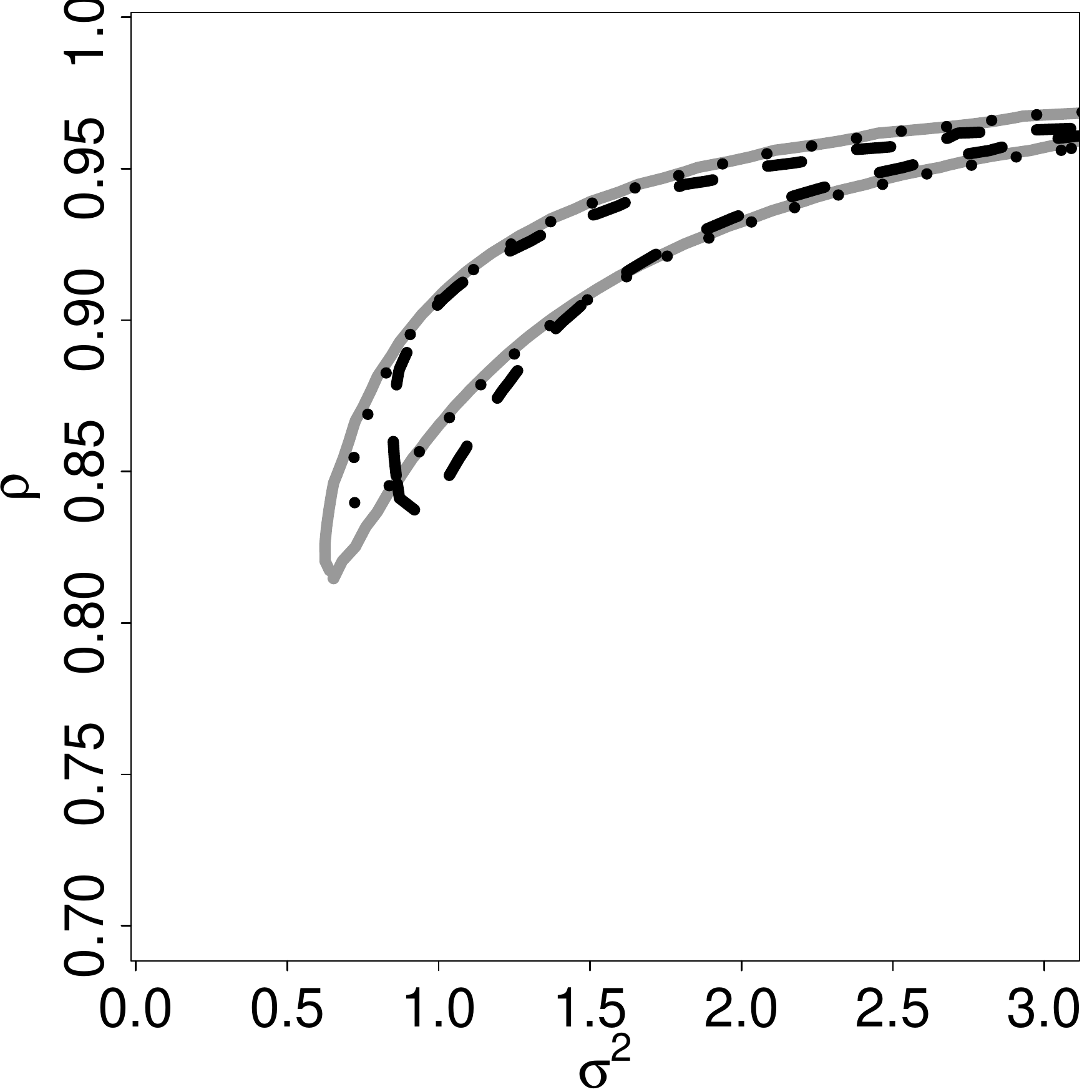}\vspace{.2cm}\\
	\quad\;{\bf {\small(d)} } & \quad\;{\bf {\small(e)} } & \quad\;{\bf {\small(f)} } \vspace{.2cm}\\
	\includegraphics[scale=0.27]{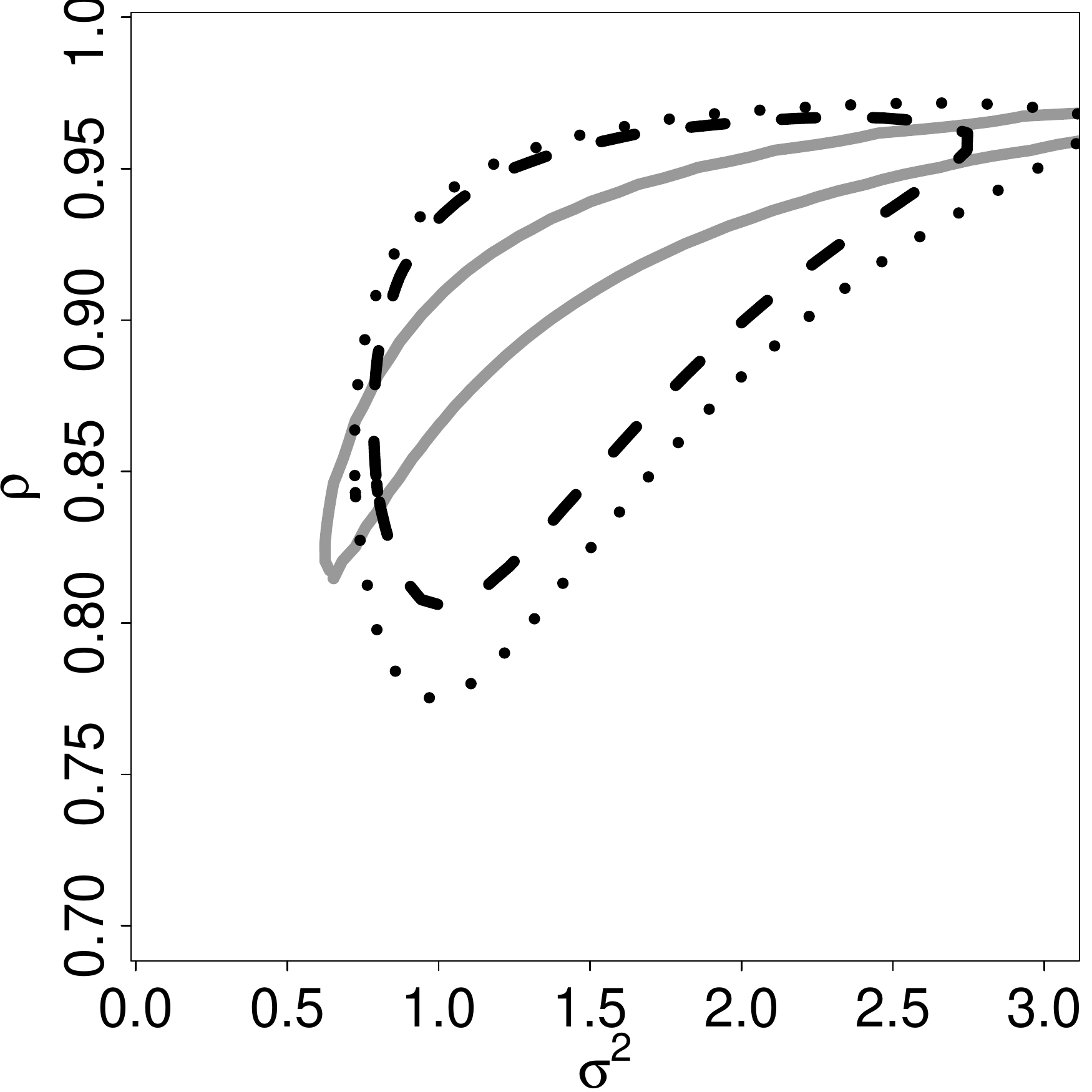}& \includegraphics[scale=0.27]{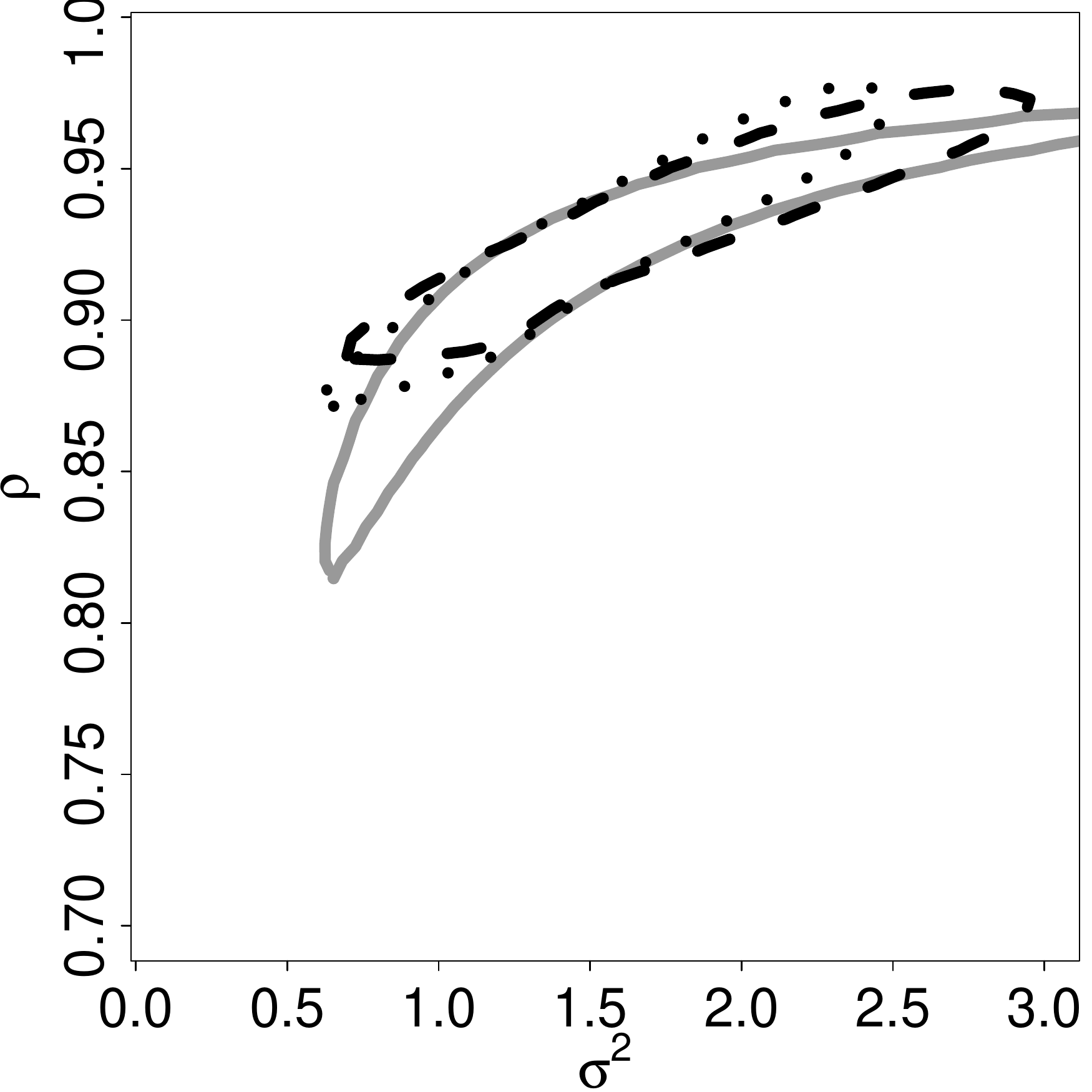} & \includegraphics[scale=0.27]{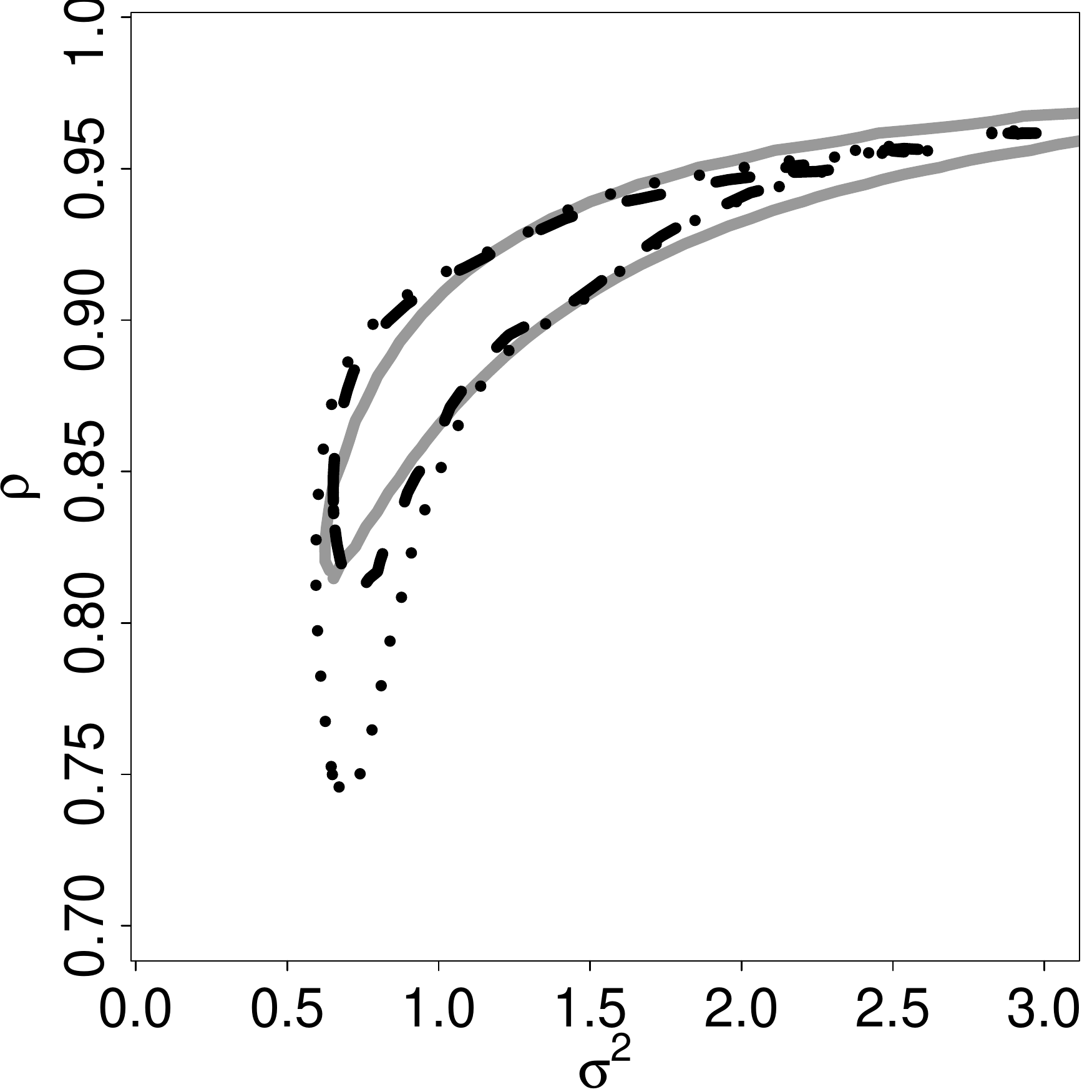}
	\end{tabular}
\end{center}
\caption{Multivariate normal model: confidence regions for $(\sigma^2,\rho)$ with nominal level $1-\alpha=0.95$, with known $\mu=0$ from a simulated sample with $n=10$ and $q=30$. In each plot confidence regions in gray solid line is obtained from $w(\theta)$. Confidence regions in dashed and dotted lines derive from pairwise likelihood statistics computed by using $J(\hat\theta_p)$ and $H(\hat\theta_p)$ and $\hat J(\hat\theta_p)$ and $\hat H(\hat\theta_p)$, respectively. In particular: (a) $pw^{*}_{sp}(\theta)$; (b) $pw_w(\theta)$, $pw^e_w(\theta)$; (c) $pw_s(\theta)$, $pw_s^e(\theta)$; (d) $pw_1(\theta)$, $pw_1^e(\theta)$; (e) $pw_{cb}(\theta)$, $pw_{cb}^e(\theta)$; (f) $pw_{inv}(\theta)$, $pw^e_{inv}(\theta)$\label{cr1}}
\end{figure}

In order to investigate the reliability of the proposed test and the ones based on pairwise likelihood statistics, it is useful to analyze the shape of the associated confidence sets and to compare them with the one provided by the full log-likelihood ratio. In Fig.~\ref{cr1} we display confidence sets for $(\sigma^2,\rho)$ with nominal level $1-\alpha=0.95$, based on statistics of Table~\ref{tab:TAB1}, from a simulated sample with $n=10$, $q=30$, $\mu=0$, $\sigma^2=1$, and $\rho=0.9$. For this analysis, the location parameter $\mu$ is considered as known. Although all confidence sets cover the true parameter value, the ones provided by $pw_1(\theta)$, $pw_w(\theta)$, and $pw_{cb}(\theta)$ depart remarkably from that of $w(\theta)$. In particular, $pw_1(\theta)$ generates a confidence set that is quite inflated and almost includes the one of $w(\theta)$, whereas Wald-type confidence sets are narrow and elliptically shaped. On the other hand, confidence sets provided by $pw_{sp}(\theta)$, $pw_s(\theta)$, and $pw_{inv}(\theta)$ resemble the gold standard. It is also worth to note how the shape of confidence sets derived from pairwise likelihood statistics is affected by the use of $J(\theta),\, H(\theta)$ and $\hat J(\theta),\, \hat H(\theta)$.

\begin{table}[!h]
\caption{Multivariate normal model: empirical coverage probabilities of three dimensional confidence regions for $\theta=(\mu,\sigma^2,\rho)$. The superscript $e$ refers to statistics computed  by using the elements of the expected Godambe information.}
\begin{center}
\small{
\begin{tabular}{l|ccc|ccc|ccc}\toprule
	& \multicolumn{3}{c}{$\rho=0.2$}&\multicolumn{3}{c}{$\rho=0.5$}&\multicolumn{3}{c}{$\rho=0.9$}\\ \hline
 $1-\alpha$ & 0.90 & 0.95 & 0.99 & 0.90 & 0.95 & 0.99 & 0.90 & 0.95 & 0.99 \\ 
  \hline
  $w(\theta)$ & 0.8802 & 0.9375 & 0.9858 & 0.8795 & 0.9367 & 0.9858 & 0.8800 & 0.9365 & 0.9859 \\ 
  $pw_{sp}^{*}(\theta)$ & 0.8644 & 0.9282 & 0.9820 & 0.8722 & 0.9300 & 0.9833 & 0.8650 & 0.9254 & 0.9809 \\ 
  $pw_w(\theta)$ & 0.5215 & 0.5855 & 0.6842 & 0.3273 & 0.3733 & 0.4567 & 0.1280 & 0.1466 & 0.1815 \\ 
  $pw_s(\theta)$ & 0.7733 & 0.8826 & 1.0000 & 0.7727 & 0.8826 & 1.0000 & 0.7747 & 0.8826 & 1.0000 \\ 
  $pw_1(\theta)$ & 0.7847 & 0.8442 & 0.9194 & 0.7505 & 0.8179 & 0.9058 & 0.7540 & 0.7823 & 0.8197 \\ 
  $pw_{cb}(\theta)$ & 0.5570 & 0.6250 & 0.7286 & 0.4201 & 0.4829 & 0.5906 & 0.1689 & 0.1991 & 0.2581 \\ 
  $pw_{inv}(\theta)$ & 0.7955 & 0.8950 & 0.9786 & 0.7980 & 0.8791 & 0.9516 & 0.9122 & 0.9462 & 0.9758 \\ \midrule

  $pw^e_w(\theta)$ & 0.7618 & 0.8155 & 0.8840 & 0.7286 & 0.7853 & 0.8601 & 0.5758 & 0.6194 & 0.6865 \\ 
  $pw^e_s(\theta)$ & 0.9051 & 0.9443 & 0.9805 & 0.9038 & 0.9435 & 0.9807 & 0.9040 & 0.9433 & 0.9807 \\ 
  $pw^e_1(\theta)$ & 0.8133 & 0.8673 & 0.9336 & 0.8136 & 0.8692 & 0.9361 & 0.8407 & 0.8983 & 0.9613 \\ 
  $pw^e_{cb}(\theta)$ & 0.7885 & 0.8459 & 0.9126 & 0.7858 & 0.8463 & 0.9190 & 0.6296 & 0.6836 & 0.7610 \\ 
  $pw^e_{inv}(\theta)$ & 0.9080 & 0.9528 & 0.9883 & 0.8940 & 0.9477 & 0.9889 & 0.8699 & 0.9276 & 0.9802 \\ 
   \bottomrule
\end{tabular}
}

\label{tab:TAB1}
\end{center}
\end{table}

\subsection{Robust First Order Autoregression}\label{Ex_AR}
We consider a stationary process $\left\{Y_j\right\}_{j\in \mathbbm Z}$, modeled as a first order autoregressive model 
\begin{equation}
\label{AR1}
Y_j=\phi_0+\phi_1 Y_{j-1}+\epsilon_j, 
\end{equation}
$\phi_0\in \mathbbm R,\,\phi_1\in(-1,1)$ and $\epsilon_j$ independent and normally distributed with mean 0 and variance $\sigma^2$. Under these assumptions any trajectory of length $q$ can be thought of as a normal random vector with expectation $(\phi_0/(1-\phi_1),\dots,\phi_0/(1-\phi_1))^\T\in \mathbbm R^q$ and covariance matrix $\Sigma$ having generic element $\Sigma_{jk}=\sigma^2\phi_1^{|j-k|}/(1-\phi_1^2),\,j,k=1,\dots,q$. 

Instead of considering bivariate marginal distributions for pairs of contiguous observations \citep{PSS}, the pairwise log-likelihood function for $\theta=(\phi_0,\phi_1, \sigma^2)$ is derived here by means of univariate conditional distributions $Y_j|Y_{j-1}=y_{j-1} \sim N(\phi_0+\phi_1y_{j-1},\sigma^2)$, and is:
\begin{equation}
\label{PL_ex2}
pl(\theta)=-\frac{(q-1)}{2}\log\sigma^2 - \frac{1}{2\sigma^2}\sum_{r=2}^q\left( y_r-\phi_0-\phi_1y_{r-1} \right)^2.
\end{equation}
The resulting pairwise score function leads to the ordinary least squares estimate of $\theta$ that can be easily robustified by using a Mallows-type estimate for $\phi_0$ and $\phi_1$ and Huber's Proposal 2 for $\sigma^2$. This is obtained by solving the system of estimating equations
\begin{equation}
\begin{array}{l}
\displaystyle \sum_{j=2}^q \psi_a(r_j) = 0\\
\displaystyle \sum_{j=2}^q \psi_a(r_j)\psi_b(y_{j-1}) = 0\\
\displaystyle \sum_{j=2}^q \psi_c(r_j)^2 - (q-1)\beta(c) = 0,
\end{array} 
\label{eq_AR}
\end{equation}
where $r_j=\left(y_j-\phi_0-\phi_1y_{j-1}\right)/\sigma$, $\psi_k(r)=\min\left\{k,\max(-k,r)\right\},\,k>0$ and $\beta(k)$ is a factor to ensure consistency at the model; see  \cite{Huber1981}, \cite{HuberRonchetti2009}. 

In order to consider both contaminated and non-contaminated series, we included an additive outlier term
in \eqref{AR1}, that becomes:
\begin{equation}
\label{outlier}
	Y_j=\phi_0 + \phi_1Y_{j-1} + \epsilon_j + u_j,
\end{equation}
where $u_j\sim(1-\xi)\delta_0+\xi N(\mu_u,\sigma^2_u)$, $\xi\in[0,1]$ and $\delta_0$ is a point mass distribution located at zero.

We performed the simulation study by drawing 100000 series of length $q=50$ from model \eqref{outlier}.
We set the true parameter value to have components $\phi_0=0$, $\sigma^2=1$, and $\phi_1=\left\{0.2, 0.5, 0.9\right\}$ and we generated contaminated series by letting $\xi=0.05$, $\mu_u=\phi_0/(1-\phi_1)$ and $\sigma^2_u=25\sigma^2.$ $\xi=0$
corresponds to the case of non-contaminated series. 
For each replication we computed the nonparametric saddlepoint test statistic as well as its bootstrap version using the estimating equations in $\eqref{eq_AR}$. They are denoted by $pw_{sp}(\theta;\gamma)$ and $pw_{sp}^{*}(\theta;\gamma)$ respectively, with $\gamma=(a, b, c)$. 
The choice $\gamma_1=(1.3, 1.3, 1.3)$ gives a bounded estimating function and
leads to a robust estimator with high efficiency at the normal model. The choice
$\gamma_2=(\infty,\infty,\infty)$ defines the classical unbounded
estimating function and leads to a non-robust estimator.

It is worth noticing that in order to preserve the dependence structure of the series and to be consistent with the specification of \eqref{AR1}, pairs of data points $(y_{j-1},y_j)$ must be resampled instead of single observations $y_j$ for the evaluation of  $pw_{sp}^{*}(\theta;\gamma).$ 

In Table~\ref{tab:TAB2} we report empirical coverage probabilities of confidence regions for $\theta$. When $\xi=0$, the comparison between $pw_{sp}(\theta;\gamma_1)$ and $pw_{sp}(\theta;\gamma_2)$ shows that the use of a bounded estimating function speeds up the convergence to the $\chi^2$ distribution. Moreover, empirical coverages of  $pw_{sp}(\theta;\gamma_1)$ and $pw_{sp}^{*}(\theta;\gamma_1)$ are very close and their accuracy is comparable to the one of the full log-likelihood ratio $w(\theta)$. When contamination occurs, the coverage levels of nonparametric saddlepoint test statistics, computed with a bounded estimating function, remain quite stable, while those of the log-likelihood ratio and $pw_{sp}(\theta;\gamma_2)$ drop away, as one would expect. 

\begin{figure}[!h]
\begin{center}
	\begin{tabular}{cc}
	\quad\;{\bf {\small$\xi=0$} } & \quad\;{\bf {\small$\xi=0.05$} }\\
	\includegraphics[height=.28\textheight, width=.48\textwidth]{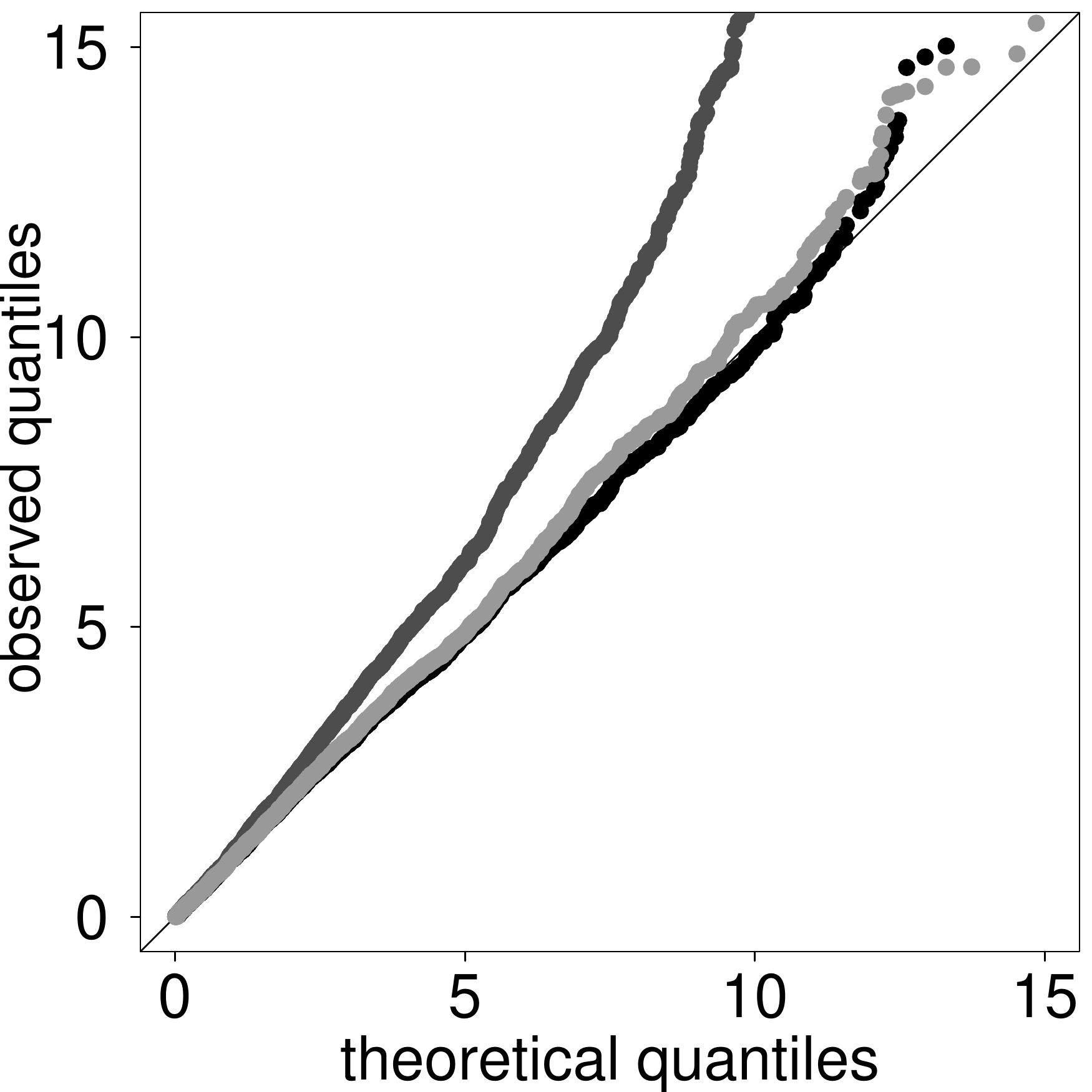} & \includegraphics[height=.28\textheight, width=.48\textwidth]{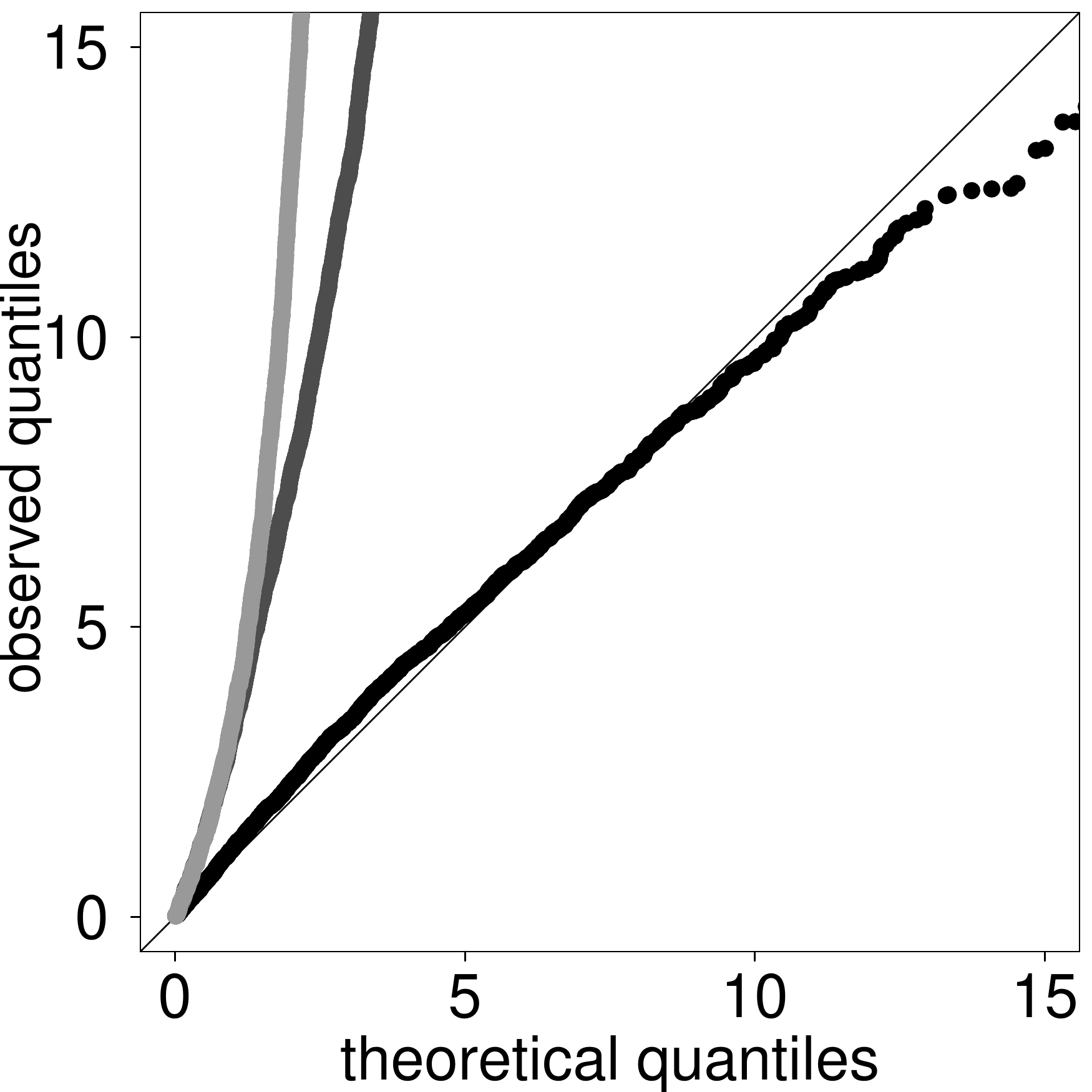} \vspace{.2cm}\\
	\end{tabular}
\end{center}
\caption{First order autoregressive model: Q-Q plots for some statistics against theoretical quantiles of the $\chi^2_3$. In black $pw_{sp}(\theta;\gamma_1)$, in dark grey $pw_{sp}(\theta;\gamma_2)$, and in light grey $w(\theta)$}
\label{qqAR}
\end{figure}

In Fig.~\ref{qqAR} we display Q-Q plots for some statistics in Table~\ref{tab:TAB2} when $\theta=(0, 0.5, 1)$. The $\chi^2$ approximation for $pw_{sp}(\theta;\gamma_1)$ is quite accurate, even when considering contaminated series, up to $\chi^2_{3;0.99}\approx 11$.

\begin{table}[!h]
\caption{First order autoregressive model: empirical coverage probabilities of three dimensional confidence regions for $\theta=(\phi_0,\phi_1, \sigma^2)$ by considering non-contaminated ($\xi=0$) and contaminated series ($\xi=0.05$).}
\begin{center}
\begin{tabular}{l|ccc|ccc|ccc}\toprule
&\multicolumn{3}{c}{$\phi_1=0.2$}&\multicolumn{3}{c}{$\phi_1=0.5$}&\multicolumn{3}{c}{$\phi_1=0.9$}\\\hline
  $1-\alpha$ 							   &  0.90  & 0.95   & 0.99     &  0.90  & 0.95   & 0.99  &  0.90  & 0.95   & 0.99\\ \bottomrule
&&\multicolumn{6}{c}{$\xi=0$}\\\hline
  $w(\theta)$        					    & 0.8915 & 0.9432 & 0.9876 & 0.8879 & 0.9403 & 0.9873 & 0.8478 & 0.9165 & 0.9792\\
  $pw_{sp}^{*}(\theta;\gamma_1)$        & 0.8914 & 0.9447 & 0.9892 & 0.8911 & 0.9447 & 0.9892 & 0.8911 & 0.9436 & 0.9881\\
  $pw_{sp}(\theta;\gamma_1)$            & 0.9007 & 0.9512 & 0.9885 & 0.9007 & 0.9512 & 0.9885 & 0.8946 & 0.9503 & 0.9898\\
  $pw_{sp}(\theta;\gamma_2)$            & 0.8232 & 0.8822 & 0.9534 & 0.8232 & 0.8822 & 0.9534 & 0.7764 & 0.8548 & 0.9376\\
\hline\midrule
&&\multicolumn{6}{c}{$\xi=0.05$}\\\hline
  $w(\theta)$        					    & 0.3441 & 0.3901 & 0.4641 & 0.2942 & 0.3365 & 0.4034 & 0.2315 & 0.2702 & 0.3236\\
  $pw_{sp}^{*}(\theta;\gamma_1)$        & 0.8818 & 0.9411 & 0.9877 & 0.8918 & 0.9456 & 0.9873 & 0.8902 & 0.9422 & 0.9869\\ 
  $pw_{sp}(\theta;\gamma_1)$            & 0.8921 & 0.9517 & 0.9917 & 0.8976 & 0.9508 & 0.9907 & 0.8728 & 0.9410 & 0.9915\\ 
  $pw_{sp}(\theta;\gamma_2)$            & 0.4612 & 0.5413 & 0.6599 & 0.3591 & 0.4328 & 0.5608 & 0.2659 & 0.3215 & 0.4251\\ 
\bottomrule
\end{tabular}
\end{center}
\label{tab:TAB2}
\end{table}

\subsection{Geostatiscal model}
Let $\left\{ Y(s), s=(s_1,\dots,s_q)\right\}$, be a stationary Gaussian random field with zero mean and exponential covariogram 
\begin{displaymath}
	\text{cov}\left[Y(s_j), Y(s_k); \theta \right]=\sigma^2\exp\left(-3||h_{jk}||/\phi\right)=\sigma^2\rho_{jk}(\phi)
\end{displaymath}
where, $h_{jk}=(s_j-s_k)$, $j,k=1,\dots,q$, $\theta=(\sigma^2,\phi)$, $||\cdot||$ is the Euclidean norm. The process is supposed to be observed on a regular lattice and we assume that the sites $s_j's$ are coordinates in $\mathbbm N^2$. In the following the discussion is developed in an increasing domain rather than an infill framework \citep[see, e.g.,][]{zhang05} but this choice does not affect the validity of our results.

\begin{figure}[!h]
\begin{center}
	\begin{tabular}{cc}
	\quad\;{\bf {\small$l=1$} } & \quad\;{\bf {\small$l=2$} }\\
	\includegraphics[height=.3\textheight, width=.5\textwidth]{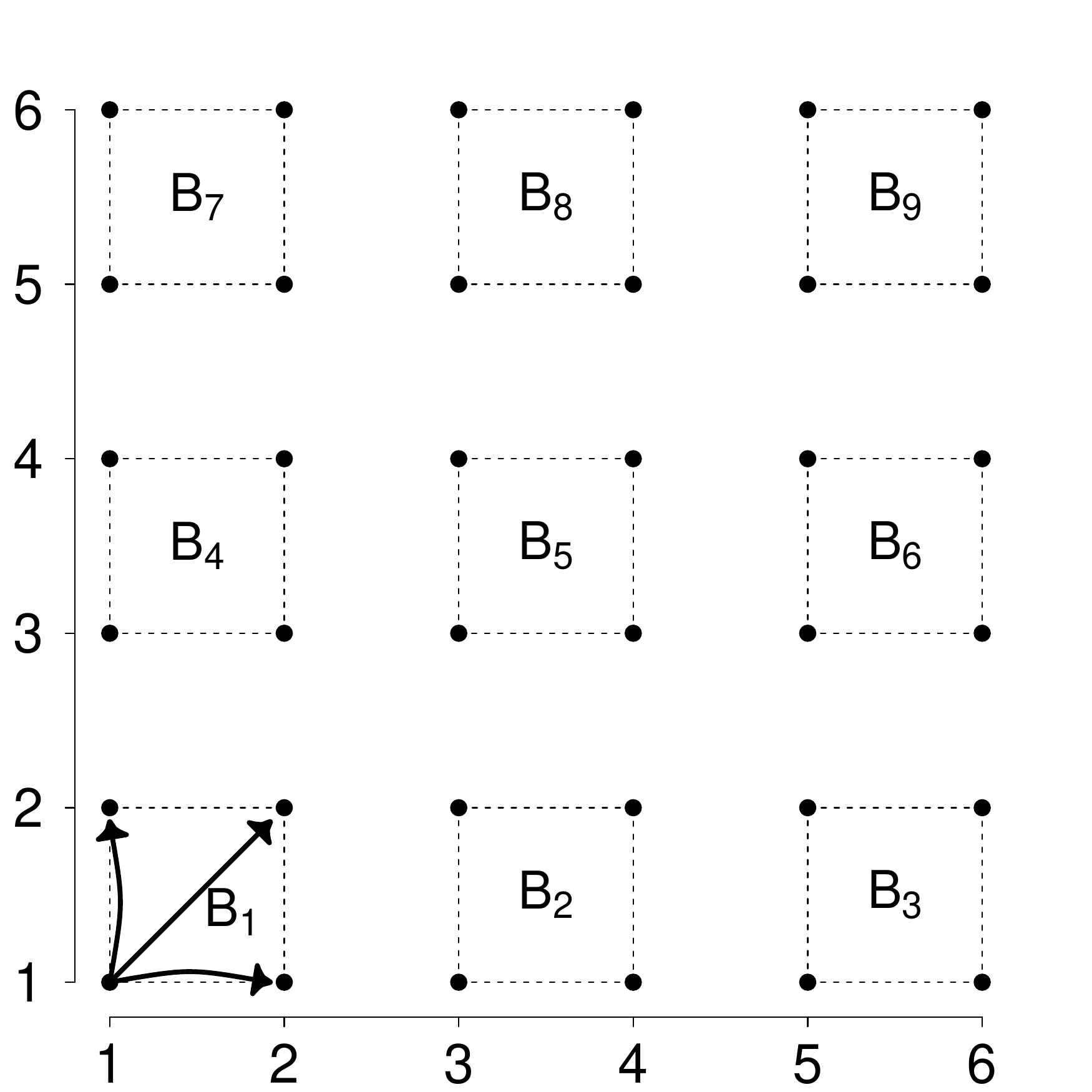} & \includegraphics[height=.3\textheight, width=.45\textwidth]{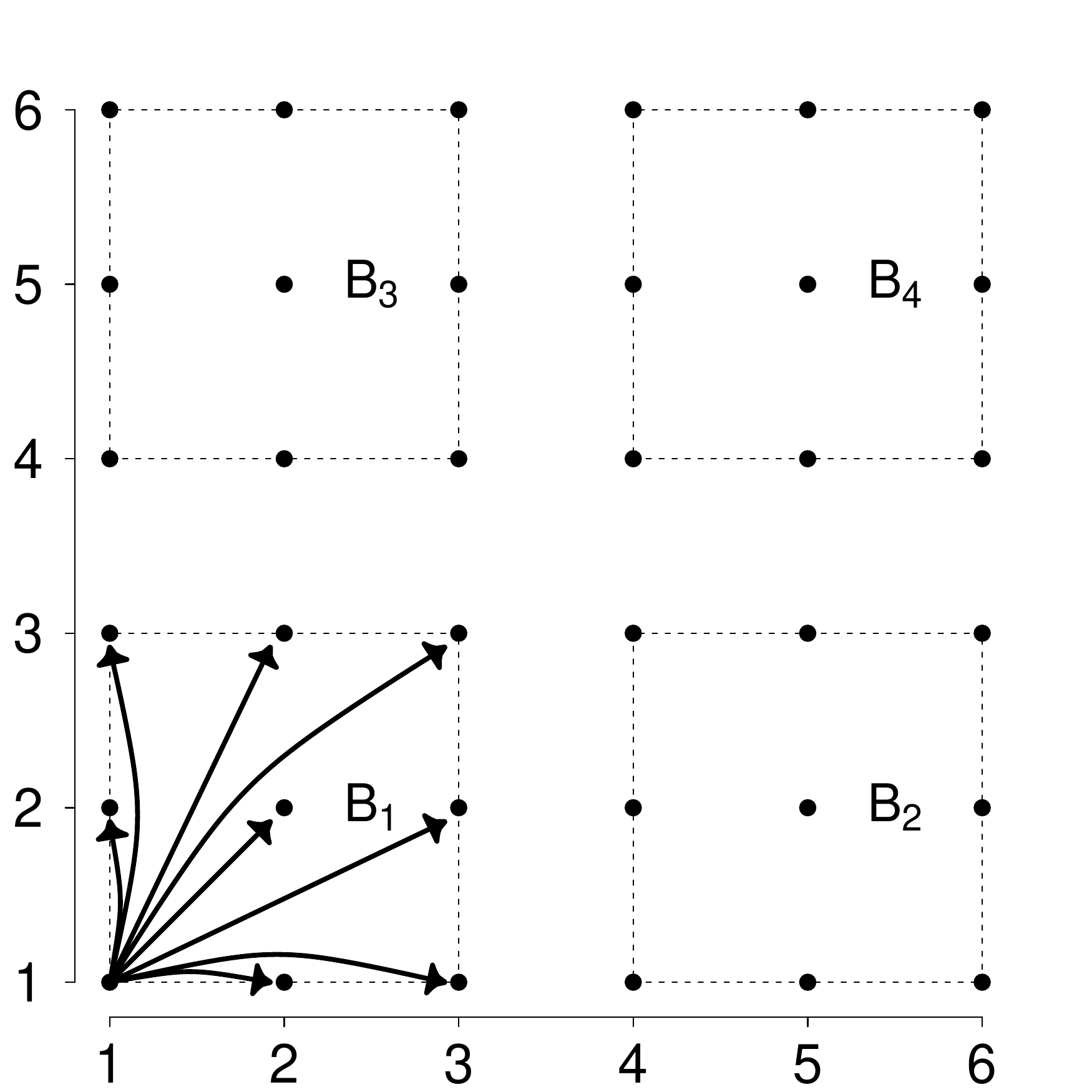} \vspace{.2cm}\\
	\end{tabular}
\end{center}
\caption{Partition of a $6\times6$ sampling region in block of observations. Dashed lines connect observations belonging to a specific block, whereas the arrows indicate which pairs are considered to compute the pairwise likelihood function}
\label{GRIDS}
\end{figure}

The pairwise log-likelihood function for $\theta$ is obtained by specifying univariate conditional distributions $Y_j|Y_k=y_k\sim N(\rho_{jk}(\phi)y_k,\sigma^2)$ and is given by
\begin{equation}
\label{pwl_spatial}
pl(\theta)=-\frac{1}{2}\sum_{j=1}^{q} \sum_{\substack{k=1\\k\neq j}}^{q}\left\{\log\sigma^2+\frac{1}{\sigma^2}\left(y_j-\rho_{jk}(\phi)y_k\right)^2 \right\}\omega(h_{jk}),
\end{equation}
where $y_j=y(s_j)$. The weights $\omega(h_{jk})$ are defined to form a disjoint partition of the sampling region in block of observations. 
Loosely speaking, the weights are chosen to form $N=[q/(1+l)]^2$ squared blocks $B_u,\,u=1,\dots,N$, each containing $(1+l)^2$ sites, where $l$ is the side length of the square.  
Inside each block only $(1+l)^2-1$ pairs are considered to compute $pl(\theta)$. Therefore, \eqref{pwl_spatial} becomes the sum of $N$ pseudo-independent blocks each of them summarizing $(1+l)^2-1$ likelihood contributions. In Fig.~\ref{GRIDS} we display how the blocks and the pairs are defined in a $6\times 6$ sampling region by considering squares with sides of length $1$ and $2$. 

It is worth to point out that the sampling region could be partitioned by constructing overlapping blocks each of them centred on a specific observation, e.g. $B_j=\left\{(y_j,y_k):\right.$\\$\left.||h_{jk}||<d\right\}$, $d>0$, $j\neq k=1,\dots,q$, and by considering different schemes to form the pairs inside each block.
For our purposes the rationale behind the splitting rule is to obtain blocks which are as uncorrelated as possible, this condition being crucial to compute both $pw_{sp}(\theta)$ and a window subsampling estimate for $J(\theta)$. 

Also in this example, $pw_{sp}(\theta)$ is computed by using a set of bounded estimating functions. From \eqref{pwl_spatial} it is easily seen that the resulting score function for a single pair is 
\begin{equation}
\begin{array}{l}
\displaystyle \ell_{\sigma^2}(\theta)=-\frac{1}{2 (\sigma^2)^2} (y_j-\rho_{jk}(\phi)y_k)^2\\
\displaystyle \ell_{\phi}(\theta)=\frac{\partial \rho_{jk}(\phi)}{\partial\phi} \frac{1}{\sigma^2} (y_j-\rho_{jk}(\phi)y_k)y_k,
\end{array} 
\label{eq_geo}
\end{equation}
which can be bounded by using the same arguments as in Example~\ref{Ex_AR}. In particular, we substitute \eqref{eq_geo} by the third and the second estimating functions in \eqref{eq_AR}, respectively.

Simulations have been run by generating $10000$ spatially correlated data from three different scenarios, corresponding to increasing levels of spatial correlation,  by setting $\sigma^2=1$ and $\phi=\left\{5,7,9\right\}$. The sampling region $\left\{1,\dots,q\right\}\times\left\{1,\dots,q\right\}$ have been increased accordingly to increasing values of $\phi$ as well as the side length of the squares defining the blocks. In particular, $q=\left\{35, 42, 54\right\}$ and $l=\left\{5,7,9\right\}$, which means setting $l$ to the effective range, i.e. the distance beyond which the correlation between pairs is less or equal to $0.05$. As a guideline we suggest to set $l$ greater or equal to the effective range, and in practical applications this can be obtained by using an empirical estimate of the correlogram.

For each replication we computed the statistics presented in Section~\ref{pairwise} as well as $pw_{sp}(\theta)$ by using the bounded counterparts of \eqref{eq_geo} with $\gamma_1=(1.3,1.3,1.3)$. The full log-likelihood ratio has not been considered in our simulations as its computation is prohibitive for the chosen values of $q$.

\begin{figure}[!h]
\begin{center}
	\begin{tabular}{cc}
	\quad\;{\bf {\small (a)} } & \quad\;{\bf {\small (b)} }\\
	\includegraphics[height=.3\textheight, width=.5\textwidth]{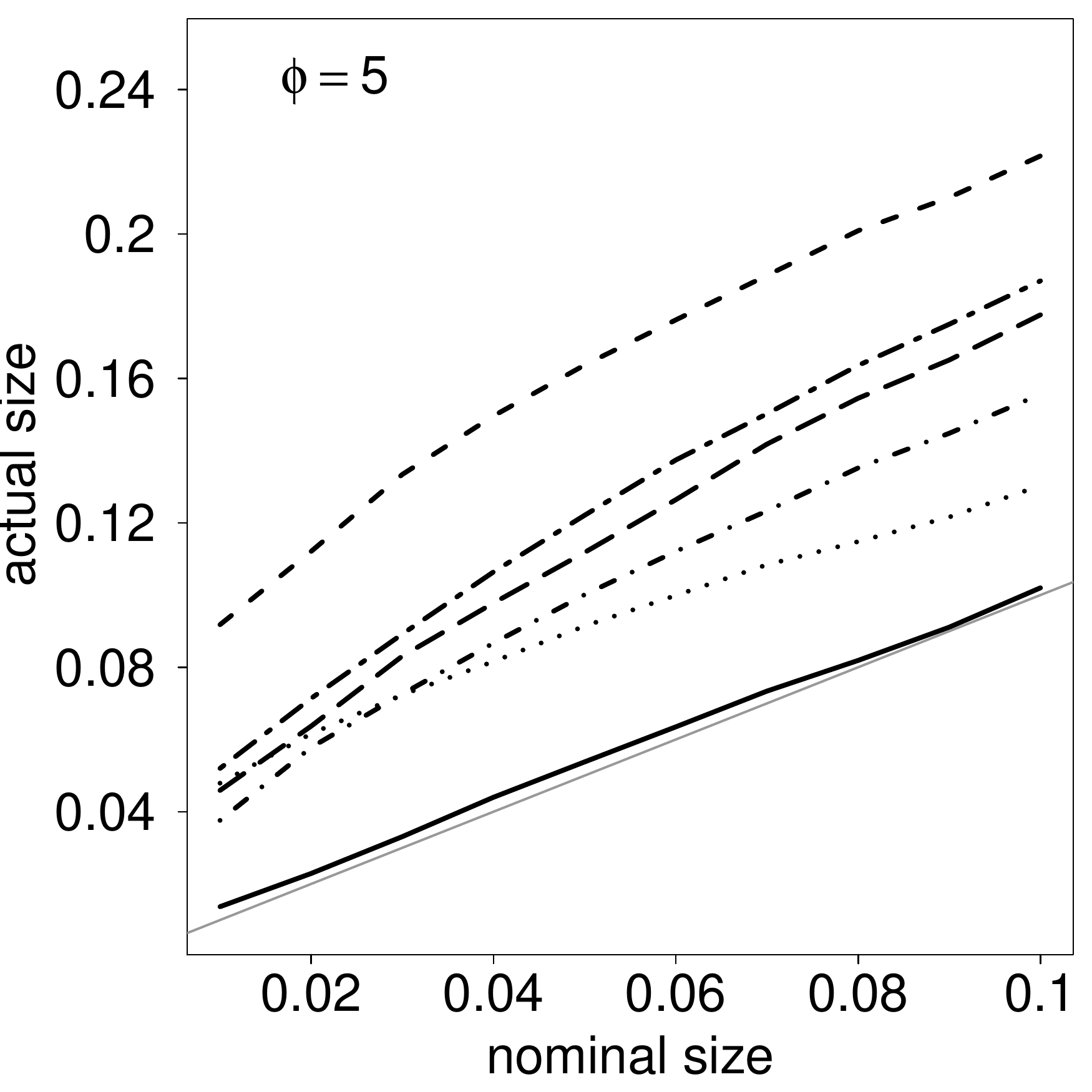} & \includegraphics[height=.3\textheight, width=.45\textwidth]{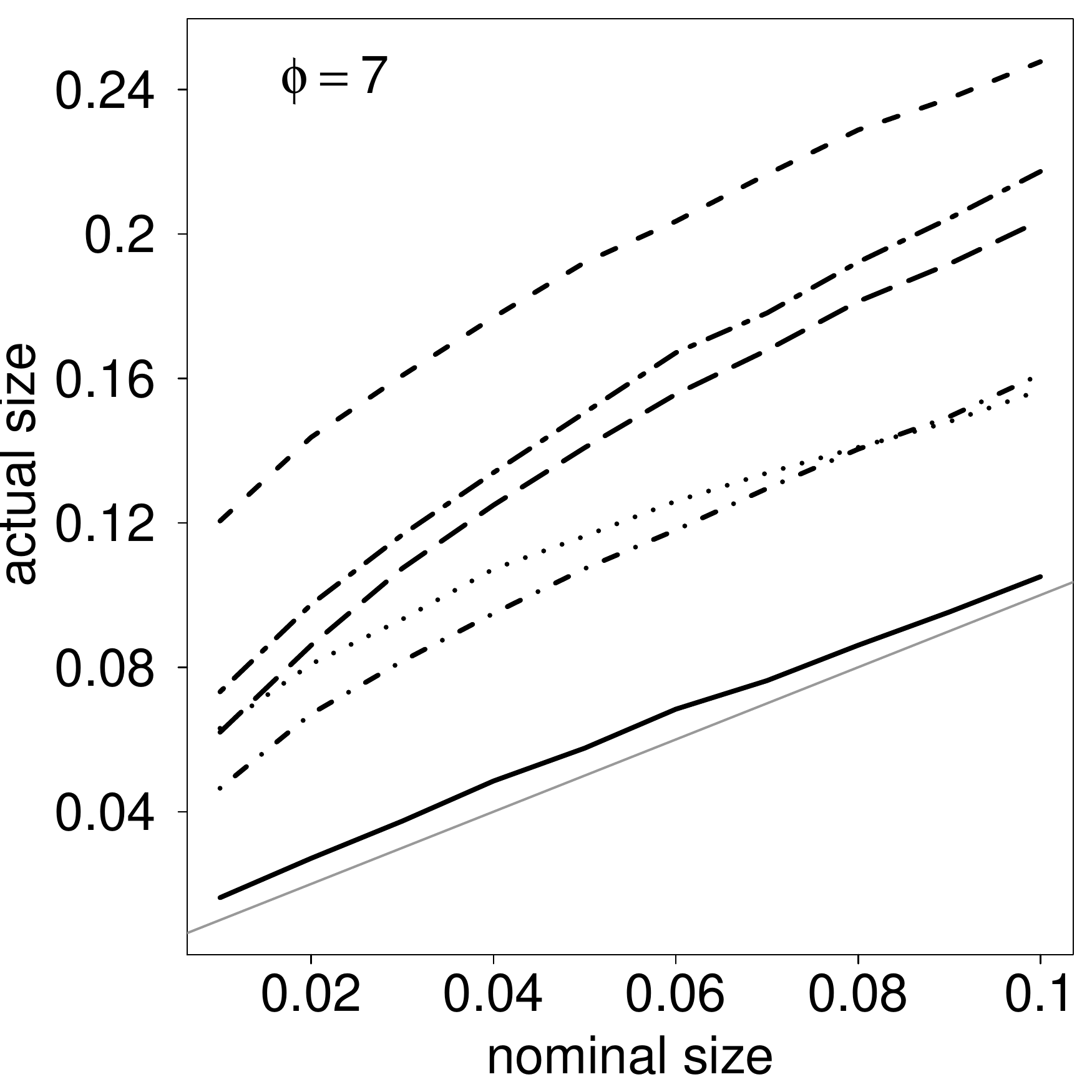} \vspace{.2cm}\\
	\quad\;{\bf {\small (c)} } & \quad\;{\bf {\small (d)} }\\
	\includegraphics[height=.3\textheight, width=.5\textwidth]{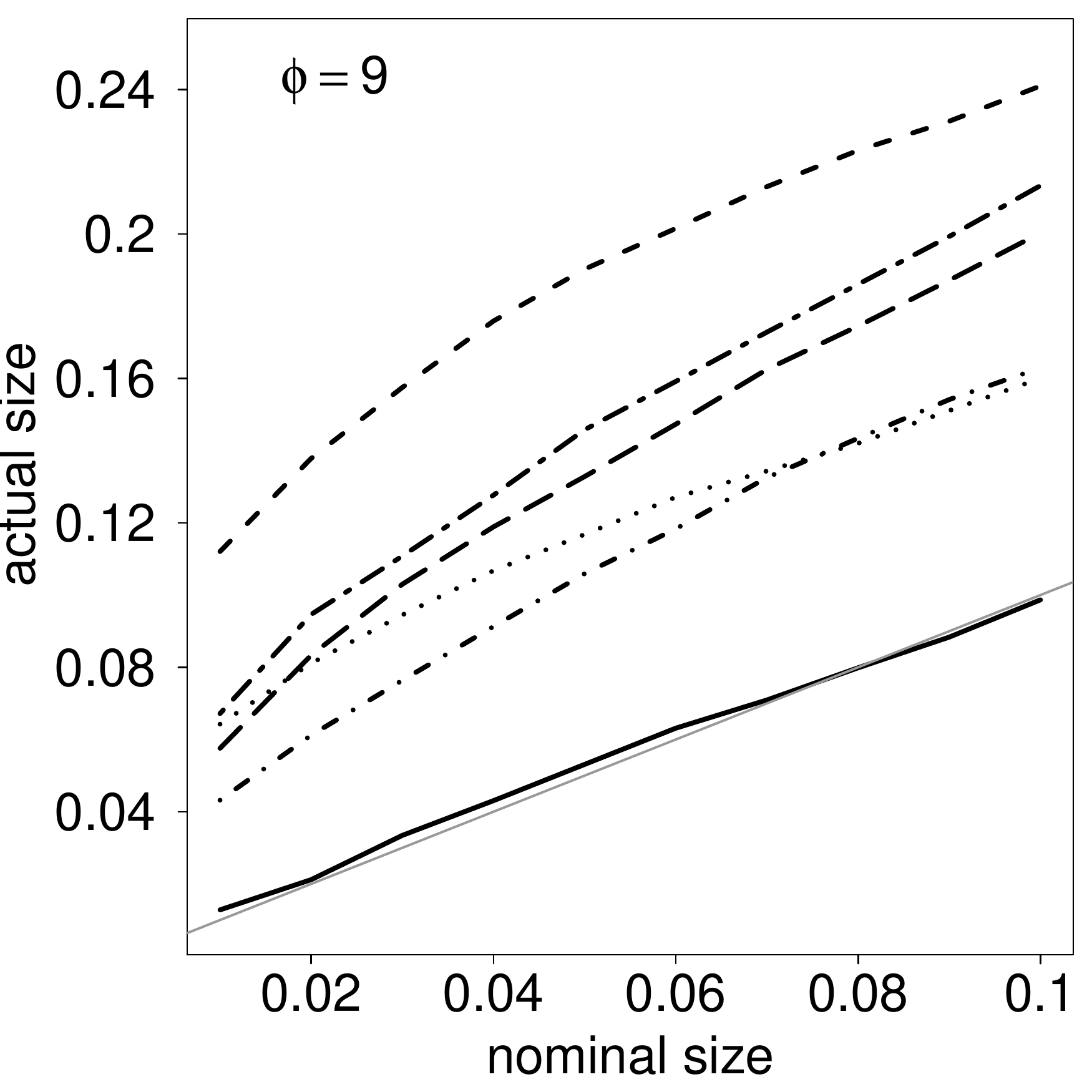} & \includegraphics[height=.3\textheight, width=.45\textwidth]{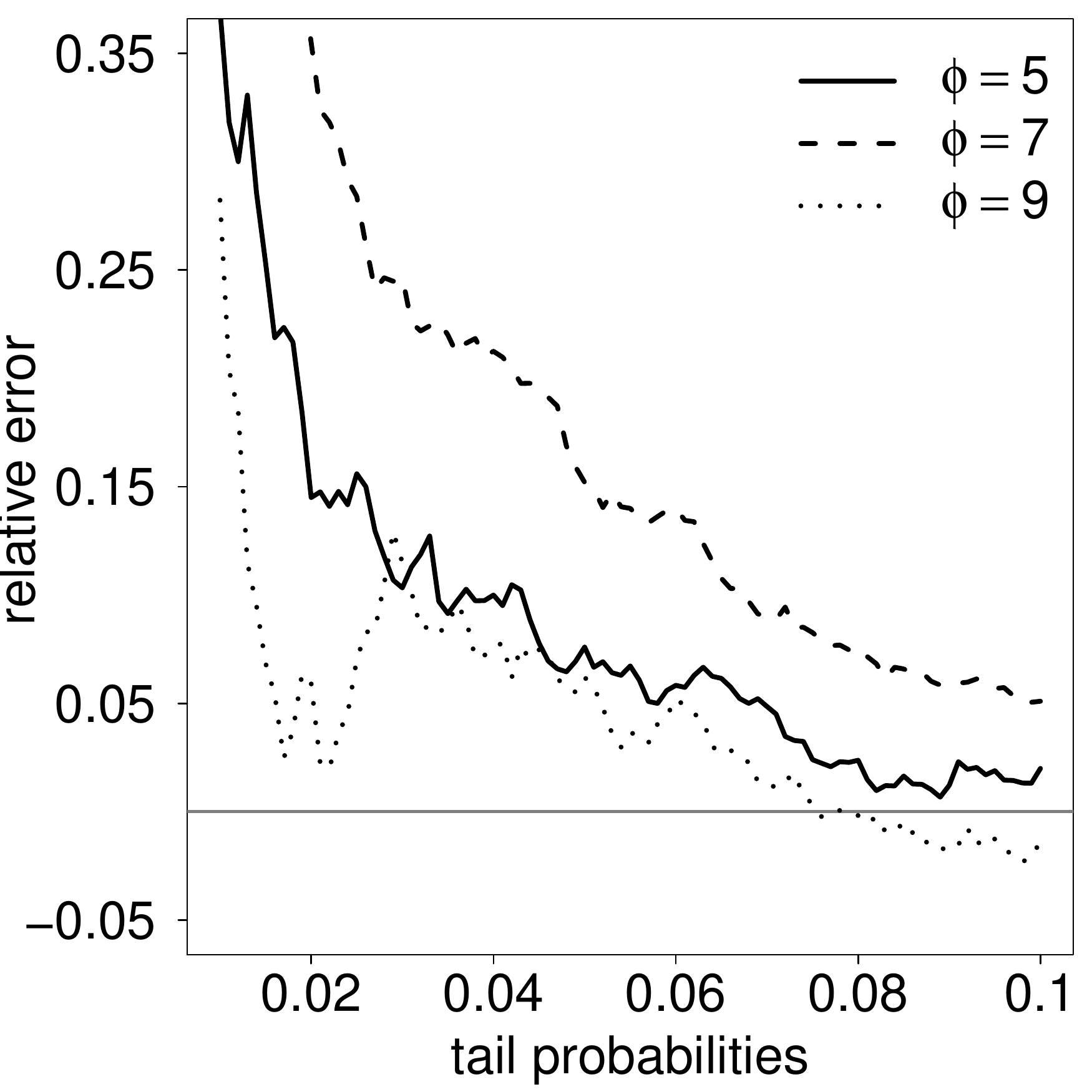} \vspace{.2cm}\\
	\end{tabular}
\end{center}
\caption{Geostatistical model: in panel (a), (b), (c) actual size is plotted against nominal size for the following test statistics: $(\solidrule[.3cm])\, pw_{sp}(\theta;\gamma)$, $(\dashedrule) \,pw_w(\theta)$, $(\dottedrule) \,pw_s(\theta)$, $(\dotteddashrule) \,pw_1(\theta)$, $(\longdashedrule)\, pw_{cb}(\theta)$, $(\shortlongdashedrule)\,pw_{inv}(\theta)$. In panel (d) approximation of the relative error for tail area probabilities provided by $pw_{sp}(\theta;\gamma)$}
\label{SP_PLOTS}
\end{figure}

In Fig.~\ref{SP_PLOTS}(a, b, c) we plot the actual sizes against the nominal sizes of tests for the three settings considered. Overall, the actual distribution of $pw_{sp}(\theta;\gamma_1)$ is closer to the $\chi^2_2$ than the ones of the other statistics. 
In panel (d) of Fig.~\ref{SP_PLOTS} we display the relative error for the tail area probabilities defined as $(P\left[pw_{sp}(\theta;\gamma)\geq\chi^2_{2;1-\alpha}\right]-\alpha)/\alpha$, for $\alpha\in\,(0.01, 0.1)$. The plot confirms that the approximation is quite accurate uniformly regardless the strength of the spatial dependence.

\section{Concluding Remarks}\label{conclusions}
We introduced in the pairwise likelihood framework a second-order accurate test statistic derived by using saddlepoint techniques. 
The new test is appealing as it circumvent the specification of the joint density and only requires the availability of the pairwise score function.
Moreover, it exhibits several desirable properties which are not shared by
the available tests. In particular, it does nor require the availability of
the Godambe information matrix of the full model, which is the case for other
standard tests. This opens up the actual possibility to
perform small sample asymptotics's inference in rather complex, yet little explored, frameworks.

\section*{Acknowledgements}
The authors would like to thank L. Pace for helpful comments.

\section*{Appendix}
{\bf Conditions}
\begin{description}
\item[(A.1):] $H(\theta)$ is continuous in $\theta$ and $|H(\theta_0)|\neq 0$;
\item[(A.2):] The components in $ps(\theta;y)$ as well as their first four derivatives with respect to $\theta$ 
exists and are bounded and continuous; 
\item[(A.3):] The cumulant generating function of $ps(\theta;Y)$ exists and the distribution function of the random vector $U=(ps(\theta; Y),S(\theta),Q(\theta))$ admits an Edgeworth expansion, where $S(\theta)$ is formed by the elements of 
$ps(\theta;Y)ps(\theta;Y)^\T$ and $\partial ps(\theta;Y)/\partial\theta^\T$, whereas $Q(\theta)$ has components 
$\partial S(\theta)/\partial\theta^\T$.
\end{description}
Condition (A.1) essentially ensures that there exists a compact subset of $\mathbbm R^p$, $\theta_0$ being an interior point of it, in which $\theta_0$ is the unique solution to $\mathbbm E[ps(\theta)]=0$. Concerning condition (A.3), the reader may refer to \citet{FieldRobinsonRonchetti2008} for a detailed account of
this technical condition.

{\bf Proof of Theorem}.
Let $y^{*}$ be a bootstrap version of $y$ obtained by sampling according to the set of probabilities $\left\{w_i(\theta_0)\right\}$, $\hat\theta^{*}_p$ be the solution to $\sum w_i(\theta_0)ps(\theta;y_i^{*})=0$, and finally denote by $P_{w}[\cdot]$ the probability under the discrete distribution defined by $\left\{w_i(\theta_0)\right\}$. 
The proof proceeds along the lines of that of Theorem $1$ in 
\cite{MaRonchetti2011} and is splitted into two steps: first the size 
of the error of the bootstrap $p$-value $P_w[pw^{*}_{sp}(\theta_0)\geq pw_{sp}(\theta_0)^{obs}]$ is established, then it is linked to the $p$-value $P[pw_{sp}(\theta_0)\geq pw_{sp}(\theta_0)^{obs}]$. 

From \cite{RobinsonRonchettiYoung2003} we have
\begin{displaymath}
P_w[pw^{*}_{sp}(\theta_0)\geq pw_{sp}(\theta_0)^{obs}]=[1-Q_p(pw_{sp}(\theta_0)^{obs})](1+O(n^{-1})),
\end{displaymath}
and from this relation it is easily seen that bootstraping the proposed statistic according to $\left\{w_i(\theta_0)\right\}$ leads to a $p$-value which error size is relative and of second-order. Then, from the results in \citet{FieldRobinsonRonchetti2008} about second-order bootstrap tests, we
obtain
\begin{eqnarray*}
P_{H_0}[pw_{sp}(\theta_0)\geq pw_{sp}(\theta_0)^{obs}]&=&P_w[pw^{*}_{sp}(\theta_0)\geq pw_{sp}(\theta_0)^{obs}](1+O(n^{-1}))\\
															 &=&[1-Q_p(pw_{sp}(\theta_0)^{obs})](1+O(n^{-1})),
\end{eqnarray*}
and this proves the theorem.

\end{document}